\documentclass[10pt, a4paper,onecolumn]{article} 
\newcommand{\bra}[1]{\langle #1|}
\newcommand{\ket}[1]{|#1\rangle}
\newcommand{\expt}[1]{\langle#1\rangle}

\bibliography{mylib}

\title{The Quantum Bayes Rule and Generalizations from the Quantum Maximum Entropy Method}
\author{Kevin Vanslette\\ kvanslette@albany.edu\\
\\Department of Physics, University at Albany (SUNY)\\
Albany, NY 12222, USA}

\date{\today} 

\begin{document}

\maketitle
\abstract{The recent article ``Entropic Updating of Probability and Density Matrices" \cite{QRE} derives and demonstrates the inferential origins of both the standard and quantum relative entropies in unison. Operationally, the standard and quantum relative entropies are shown to be designed for the purpose of inferentially updating probability distributions and density matrices, respectively, when faced with incomplete information. We call the inferential updating procedure for density matrices the ``quantum maximum entropy method". Standard inference techniques in probability theory can be criticized for lacking concrete physical consequences in physics; but here, because we are updating quantum mechanical density matrices, the quantum maximum entropy method has direct physical and experimental consequences.  The present article gives a new derivation of the Quantum Bayes Rule, and some generalizations, using the quantum maximum entropy method while discuss some of the limitations the quantum maximum entropy method puts on the measurement process in Quantum Mechanics.  }

\section{Introduction}
The recent article ``Entropic Updating of Probability and Density Matrices" \cite{QRE} derives and demonstrates the inferential origins of both the standard and quantum relative entropies in unison. The derivations of the standard and quantum relative entropies in \cite{QRE} were not rudimentary; rather, a set of inferentially guided \emph{design criteria} were proposed to \emph{design} a function capable of accurately updating probability distributions when faced with incomplete information. The solution has the functional form of the standard relative entropy and thus the standard relative entropy is the functional designed for the purpose of probability updating.  Similar (design) derivations exist \cite{book,Shore1,Shore2,Csiszar1991,Skilling1,Skilling2,Skilling3}, but the number of required design criteria was reduced in \cite{QRE}. What is particularly pleasant in \cite{QRE} is the equal implementation of the same design criteria to design a functional capable of updating density matrices. This parallel derivation shows the quantum relative entropy is designed to update density matrices -- formulating an inferentially oriented quantum maximum entropy method. Not only was the quantum relative entropy found in \cite{QRE}, but we also learned how to use it. This discussion saturates the previous article. Here we provide a new derivation of the Quantum Bayes Rule (QBR), discuss the physical implications entropic methods puts on the measurement process in Quantum Mechanics (QM), and briefly discuss how the quantum maximum entropy method provides some simple generalizations of the QBR.  

In QM the wavefunction has two modes of evolution \cite{Neumann,Luders}: one is the continuous unitary evolution given by the dynamical Schr\"{o}dinger equation, while the other is the discrete collapse of the wavefunction that occurs when a detection is made. The collapse postulate is generally implemented \emph{ad hoc} to empirically represent the effect of detection, and more recently the Quantum Bayes Rule \cite{Korotkov1,Korotkov2,Jordan} (also known as the fundamental theorem of quantum measurement \cite{Jacobs} or the positive operator-valued measure (POVM) formalism \cite{Davies,Kraus,Holevo,Nielsen}),
\begin{eqnarray}
\hat{\rho}_{\theta}=\frac{A_{x}\hat{\varphi}_{\theta}A^{\dag}_{x}}{\mbox{Tr}(A_{x}\hat{\varphi}_{\theta}A^{\dag}_{x})},\label{Qbayes}
\end{eqnarray}
quantifies collapse under a POVM measurement, $A^{\dag}_{x}A_{x}$, where $\sum_x A^{\dag}_{x}A_{x}=\hat{1}$, which is a generalization of L\"{u}ders Rule \cite{Luders}. Here we will derive the QBR from entropic arguments, which we claim eliminates the need for \emph{ad hoc} collapse postulates in QM. Our result further perpetuates the interpretation that entropy may be used to inferentially collapse the wavefunction using projectors \cite{Hellmann,Kostecki}, and our result is generalized to the QBR using POVM's (and a weak QBR). Rather than appealing to group theoretic arguments \cite{Hellmann,Kostecki}, our derivations are seemingly simpler as they require solving the Lagrange multiplier problem following \cite{QRE} (the derivation parallels \cite{Giffin,GiffinThesis} but using density matrices) and naturally avoids the infinite entropy problem that that appears for ``strong L\"{u}ders rule" derivation in \cite{Hellmann}. The Lagrange multiplier technique is used in the maximum entropy method community (\cite{Jaynes1,Jaynes2,Jaynesbook} and the works and conferences that have followed) for updating probability distributions, so we refer to the method of inference capable of update density matrices as the quantum maximum entropy method. 

 As both forms of the standard and quantum relative entropy resemble one another, they inevitably share analogous solutions and face similar limitations; however, because we are dealing with density matrices, these limitations have physical consequences.  In standard probability theory, there is a phrase, ``The maximum entropy method cannot fix flawed information" \cite{book}, and a similar theme permeates the inference procedure for density matrices. Because the entropy was designed to update from a prior density matrix $\hat{\varphi}$ to a posterior density matrix $\hat{\rho}$, the form of $\hat{\varphi}$ must accurately describe the prior state of knowledge of the system if $\hat{\rho}$ is going to objectively represent the updated state of knowledge for that quantum system. For instance, if our prior knowledge tells us that a particle is located within a certain interval, it makes no sense to impose that the particle have an average position anywhere but within that interval. The quantum mechanical analog of this is that a prior density matrix cannot be updated to regions not originally spanned by the prior density matrix. We derive this type of incompatibility for the quantum maximum entropy method, which we name the Prior Density Matrix Theorem (PDMT). The PDMT sheds light on some of the nontrivial notions of quantum measurement and QM in general. 

 A special case of the PDMT insists that the detection of an observable from a pure state (the collapse) is impossible without first decohering (or partially decohering) the pure state. This is a rediscovery of L\"uders' notion \cite{Luders} that the action of a measurement device is to project the pure state into a mixed state $\hat{\rho}\rightarrow \sum \hat{P}_i\hat{\rho}\hat{P}_i$,  except our argument is from purely entropic and thus inferential arguments. This concept is not as foreign or as objectionable as it may seem if we consider the well known results of the quantum two slit experiment. If a ``which slit" detection of the particle is made, then the resulting probability distribution is a decohered sum of Gaussians on the screen (after many trials), whereas omitting this detection allows for interference effects. Decoherence of the pure state was required for a which slit inference. The PDMT further insists that once the particle hits the screen, to detect its state, it must first decohere (potentially again) on the detection screen. This imprints a mixed state realization of the incoming pure or decohered state on the screen $\hat{\rho}\rightarrow \sum \hat{P}_i\hat{\rho}\hat{P}_i$, which may be detected and collapse the state. In this sense, ``collapse of the wavefunction" is better stated ``collapse of the mixed state" -- which then, as we will see, is nothing more than standard probability updating.   

In preparation for the derivation of the Quantum Bayes Rule using the quantum maximum entropy method, the derivation of Bayes Rule using the standard maximum entropy method will be reviewed \cite{QRE,Giffin}. We will introduce the PDMT and apply the quantum maximum entropy method to derive the aforementioned cases of interest. 

\section{Maximum Entropy Method}
Here we will loosely refer to $\rho(x)$ as a probability distributions rather than a probability density with the understanding that the probability of an event is actually $\rho(x)\,dx$. E.T. Jaynes is the proprietor of the maximum entropy method \cite{Jaynes1,Jaynes2,Jaynesbook}, but over the years his insights have been refined \cite{book}.  

The relative entropy and quantum relative entropy were designed for the purpose of making inference in \cite{QRE} by implementing a set of \emph{design criteria}. The design criteria are guided by the ``Principle of Minimum Updating" (PMU) -- which is the claim that a probability distribution should only be updated to the extent required by the information -- while information itself is defined operationally as being what causes probability distributions to change. This pragmatic principle enforces objectivity to this method of inference. The functional form remaining, after implementing the design criteria, takes the form of a relative entropy,
\begin{eqnarray}
S(\rho(x),\varphi(x))=-|A|\int dx\, (\rho(x)\ln\Big(\frac{\rho(x)}{\varphi(x)}\Big)-\rho(x))+C[\varphi],\label{theentropy}
\end{eqnarray}
where $C[\varphi]$ is a constant independent of $\rho$ and $|A|\neq 0$ is an arbitrary but positive constant such that we are really maximizing the entropy. Maximizing this entropy with respect to a set of expectation value constraints $\expt{A_i}=\int\,dx\,\rho(x) A_i(x)$ and normalization via the Lagrange multiplier method is setting the variation,
\begin{eqnarray}
\delta\Big(S(\rho(x),\varphi(x))-\lambda(\int\,dx\,\rho(x)-1)-\sum_i\alpha_i(\int\,dx\,\rho(x) A_i(x)-\expt{A_i})\Big)=0,
\end{eqnarray}
where \{$\lambda,\alpha_i$\} is the set of Lagrange multipliers that enforce their corresponding expectation value constraints. For arbitrary variations of $\rho(x)$,
\begin{eqnarray}
\int\,\Big(-|A|\ln\Big(\frac{\rho(x)}{\varphi(x)}\Big)-\lambda-\sum_i\alpha_iA_i(x)\Big)\delta\rho(x)\,dx=0.
\end{eqnarray}
The probability distribution that maximizes the entropy for arbitrary variations in $\rho(x)$ occurs when the terms within the parenthesis vanish; and therefore one finds canonical-like solutions,
\begin{eqnarray}
\rho(x)=\varphi(x)\exp\Big(\frac{\lambda+\sum\alpha_iA_i(x)}{|A|}\Big),
\end{eqnarray}
in general. As $|A|$ is a nonzero constant, it may be absorbed into the Lagrange multipliers, $(\lambda,\{\alpha_i\})$, so we may let it equal unity without loss of generality. Writing the normalization Lagrange multiplier in the suggestive form $Z=e^{-\lambda}$ gives,
\begin{eqnarray}
\rho(x)=\frac{\varphi(x)}{Z}\exp\Big(\sum\alpha_iA_i(x)\Big).\label{gensolrho}
\end{eqnarray}
The Lagrange multipliers are solved by evaluating their corresponding expectation value constraints, usually employing standard methods from Statistical Mechanics $\expt{A_i}=\frac{\partial}{\partial \alpha_i}\log(Z)$.  One of the design criteria (DC1') used to derive the relative entropy in \cite{QRE} enforces the Principle of Minimum Updating by stating, ``in the absence of new information, the posterior distribution $\rho$ is equal to the prior distribution $\varphi$". This is indeed the case if either no expectation value constraints are imposed, or if the imposed expectation value constraints are already satisfied by $\varphi$ -- in which case the introduced Lagrange multipliers are zero.

\paragraph{A comment on biased priors}
Entropic inference of this nature is only as useful as we are objective about our subjectivity. One should be careful not to apply nonsensical constraints, i.e. attempting to impose impossible expectation values. In such a case, the maximum entropy method provides ``no solution" to the optimization problem due to its irrationality. If a set of microstates $x\in D_0$ in a domain $D_0$ are assigned a zero prior probability $\varphi(x\in D_0|s)=0$ given some situation $s$, then it is impossible to update the posterior distribution to anything but $\rho(x\in D_0|s)=0$ for any amount of new information (as can be seen in (\ref{gensolrho})). In the same way, a delta function prior distribution $\varphi(x|s)\sim\delta(x-x_0)$, which claims complete certainty, cannot be updated. We call distributions that cannot be updated due to having poor priors ``biased" as any amount new information does not change the prescribed state of knowledge $\varphi(x|s)\rightarrow\rho(x|s)=\varphi(x|s)$. A biased state of knowledge pertaining to a situation $s$ does not imply bias for a new situation $s'$, so a realization that a nonzero (or non-infinite) probability should be assigned to the region $D_0$ admits the system is now in a new situation $s'$. An example of this from Statistical Mechanics (and also QM) occurs if the distance between the walls of an infinite potential box is enlarged such that previous zero probability regions now gain possibility. In this sense, and others, entropic updates are purely epistemic. These notions extend to density matrices as we will see later.

\paragraph{Maximum Entropy and Bayes}
  When the information provided is in the form of data, entropic updating is consistent with Bayes Rule,
\begin{eqnarray}
\rho(\theta)\stackrel{1}{=}\varphi(\theta|x)\stackrel{2}{=}\frac{\varphi(x|\theta)\varphi(\theta)}{\varphi(x)},
\end{eqnarray}
where Bayes Rule is the first equal sign and Bayes Theorem is the second equal sign \cite{book}. The leads to the realization that Bayesian and entropic inference methods are consistent with one another \cite{Giffin,GiffinThesis}. 

The posterior distribution $\rho(\theta)$ can only be realized once the data about $x$'s has been processed. This implies the state space of interest is the product space of $\mathcal{X}\times\Theta$ with a joint prior $\varphi(x,\theta)$. Suppose we collect data and observe the value $x'$. The data constrains the joint posterior distribution $\rho(x,\theta)$ to reflect the fact that the value of $x$ is known to be $x'$, that is,
\begin{eqnarray}
\rho(x)=\int d\theta \rho(x,\theta)=\delta(x-x'),\label{data}
\end{eqnarray}
however; this data constraint is not enough to specify the full joint posterior distribution, 
\begin{eqnarray}
\rho(x,\theta)=\rho(\theta|x)\rho(x)=\rho(\theta|x)\delta(x-x'),
\end{eqnarray}
because $\rho(\theta|x)$ is not determined.

 As there are many distributions that satisfy this data constraint, we rank the distributions using the relative entropy. Note that the data constraint (\ref{data}) in principle constrains \emph{each} $x$ in $\rho(x,\theta)$ so a Lagrange multiplier $\alpha(x)$ is required to tie down each  $x\in\mathcal{X}$ of the marginal distribution $\rho(x)$. Maximizing the entropy with respect to this constraint and normalization is,
\begin{eqnarray}
0=\delta\Big(S-\lambda[\int \rho(x,\theta)\,dxd\theta-1]-(\int \alpha(x)[\int\rho(x,\theta)\,d\theta-\delta(x-x')]\,dx)\Big)
\end{eqnarray}
where $\lambda$ is the Lagrange multiplier imposing normalization. This leads to the following joint posterior distribution,
\begin{eqnarray}
\rho(x,\theta)=\varphi(x,\theta)\frac{e^{\alpha(x)}}{Z}.
\end{eqnarray}
The Lagrange multiplier $Z$ is found by imposing normalization,
\begin{eqnarray}
1&=&\int\rho(x,\theta)\,dxd\theta=\frac{1}{Z}\int\varphi(x,\theta)e^{\alpha(x)} \,dxd\theta\nonumber\\
&\rightarrow&Z=\int\varphi(x,\theta)e^{\alpha(x)} \,dxd\theta.
\end{eqnarray}
The Lagrange multiplier $\alpha(x)$ is found by considering the data constraint,
\begin{eqnarray}
\delta(x-x')=\int\rho(x,\theta)d\theta=\frac{e^{\alpha(x)}}{Z}\int\,\varphi(x,\theta)d\theta=\frac{e^{\alpha(x)}}{Z}\varphi(x).
\end{eqnarray}
Substituting this solution into the joint posterior distribution gives,
\begin{eqnarray}
\rho(x,\theta)=\frac{\varphi(x,\theta)}{\varphi(x)}\delta(x-x')=\varphi(\theta|x)\delta(x-x').
\end{eqnarray}
Integrating over $x$ gives the marginalized posterior distribution,
\begin{eqnarray}
\rho(\theta)=\varphi(\theta|x')=\frac{\varphi(x'|\theta)\varphi(\theta)}{\varphi(x')},
\end{eqnarray}
which is Bayes Rule. Generalizations of Bayes Rule, such as Jeffery's Rule when the data (and constraint) is uncertain, $\int d\theta \rho(x,\theta)=\rho_D(x)$, are also consistent with the method of maximum entropy (further review can be found in \cite{ book,Giffin,GiffinThesis}). The universality of this entropic inference method is emphasized by it consistency with other forms of inference like Bayesian inference. 

\section{Quantum Maximum Entropy Method}

The derivation of the quantum relative entropy parallels the derivation of the standard relative entropy because the same \emph{design criteria} were applied in both cases, but this time to the ranking of density matrices \cite{QRE}. The form of the quantum relative entropy that saturates the \emph{design criteria} is,
\begin{eqnarray}
S(\hat{\rho},\hat{\varphi} )=-|A|\mbox{Tr}(\hat{\rho} \log \hat{\rho} -\hat{\rho}\log \hat{\varphi} -\hat{\rho})+C[\hat{\varphi}]=-|A|S_U(\hat{\rho},\hat{\varphi})+|A|\mbox{Tr}(\hat{\rho})+C[\hat{\varphi}],
\end{eqnarray}
where $S_U(\hat{\rho},\hat{\varphi})$ is Umegaki's form of the quantum relative entropy. Similarly, maximizing this entropy with respect to a set of expectation values of Hermitian operators $\{\hat{A}_i\}$, (i.e. $\mbox{Tr}(\hat{A}_i\hat{\rho})=\expt{A_i}$) and normalization is setting the variation,
\begin{eqnarray}
\delta\Big(S(\hat{\rho},\hat{\varphi})-\lambda[\mbox{Tr}(\hat{\rho})-1]-\sum_i\alpha_i[\mbox{Tr}(\hat{A}_i\hat{\rho})-\expt{A_i}]\Big)=0.
\end{eqnarray}
Arbitrary variations of $\hat{\rho}$ is,
\begin{eqnarray}
\mbox{Tr}\Big(\Big(-|A|(\log \hat{\rho} - \log \hat{\varphi})-\lambda\hat{1}-\sum_i\alpha_i\hat{A}_i\Big)\delta\hat{\rho}\Big)=0,
\end{eqnarray}
due to the cyclic property of the trace, which, after absorbing $|A|$ into the Lagrange multipliers again, gives the solution,
\begin{eqnarray}
\hat{\rho}=\exp\Big(\lambda\hat{1}+\sum_i\alpha_i\hat{A}_i+\ln(\hat{\varphi})\Big)=\frac{1}{Z}\exp\Big(\sum_i\alpha_i\hat{A}_i+\ln(\hat{\varphi})\Big),\label{rhosolution}
\end{eqnarray}
where normalization may be factored out of the exponential due to the universal commutativity of the identity operator.   The remaining problem is to solve for the $n$ Lagrange multipliers using their $n$ associated expectation value constraints. In principle their solution is found by computing $Z$ and using standard methods from Statistical Mechanics,
\begin{eqnarray}
\expt{\hat{A}_i}=\frac{\partial}{\partial \alpha_i}\ln(Z)\label{gen}
\end{eqnarray}
and inverting to find $\alpha_i=\alpha_i(\expt{\hat{A}_i})$. Using these methods, the relevant thermodynamic and quantum information theoretic results in \cite{Goold} that stem from quantum relative entropy may be reproduced and rephrased as applications of inference. Between the Zassenhaus formula
\begin{eqnarray}
e^{t(\hat{A}+\hat{B})}=e^{t\hat{A}}e^{t\hat{B}}e^{-\frac{t^2}{2}[\hat{A},\hat{B}]}e^{\frac{t^3}{6}(2[\hat{B},[\hat{A},\hat{B}]]+[\hat{A},[\hat{A},\hat{B}]])}...,
\end{eqnarray}
and Horn's inequality, in general the solutions to (\ref{gen}) lack a certain calculational elegance due to the difficulty of expressing the eigenvalues of $\hat{C}=\log(\hat{\varphi})+\sum\alpha_i\hat{A}_i$ in simple terms of the eigenvalues of the $\hat{A}_i$'s and $\hat{\varphi}$, when the matrices do not commute. The solution requires solving the eigenvalue problem for $\hat{C}$, such the the exponential of $\hat{C}$ may be taken and evaluated in terms of the eigenvalues of the $\alpha_i\hat{A}_i$'s and the prior density matrix $\hat{\varphi}$. It is at this point that the review of the relevant material has concluded.

\subsection{Prior density matrices}
If the prior density matrix $\hat{\varphi}=\hat{\varphi}^2$ is a projector, then we consider it to be a ``biased" density matrix because no amount of information can update the posterior density matrix, i.e. $\hat{\varphi}\rightarrow\hat{\rho}=\hat{\varphi}$ using entropic methods. An example using spin is discussed below to introduce the notion.

Consider the biased prior density matrix $\hat{\varphi}=\ket{+}\bra{+}$ -- the positive spin-$z$ eigenstate. To preform calculation with any rigor using this biased prior, we must unbias it slightly by considering something like $\hat{\varphi}=\lim_{\epsilon\rightarrow0}\Big((1-\epsilon)\ket{+}\bra{+}+\epsilon\ket{-}\bra{-}\Big)\equiv\lim_{\epsilon\rightarrow0}\hat{\varphi}_{\epsilon}$. We will use $\hat{\varphi}_{\epsilon}$ for the prior, and then take the limit $\epsilon\rightarrow0$ when appropriate. In attempting to force the issue, consider maximizing the relative entropy with respect to $\expt{\vec{b}\cdot\vec{\sigma}}$ such that $\hat{\rho}$ requires nonzero components along spin down $\ket{-}\bra{-}$, in contrast to $\hat{\varphi}$. Maximizing the entropy with respect to this constraint, normalization, and the biased prior gives,
\begin{eqnarray}
\hat{\rho}_{\epsilon}=\frac{1}{Z}\exp\Big(\alpha(\vec{b}\cdot\vec{\sigma})+\ln(\hat{\varphi}_{\epsilon})\Big)\equiv\frac{1}{Z}\exp(\hat{C}_{\epsilon}).
\end{eqnarray}
The Lagrange multiplier which imposes normalization may be found by diagonalizing the exponent $\hat{C}_{\epsilon}\rightarrow \hat{\Lambda}_{\epsilon}$,
\begin{eqnarray}
Z=\mbox{Tr} (\exp(\hat{C}_{\epsilon}))=\mbox{Tr} (\hat{U}_{\epsilon}\exp\Big(\hat{U}_{\epsilon}^{\dag}\hat{C}_{\epsilon}\hat{U}_{\epsilon}\Big)\hat{U}_{\epsilon}^{\dag})=\mbox{Tr}(e^{\hat{\Lambda}_{\epsilon}})=\sum_{\lambda_{\epsilon}} e^{\lambda_{\epsilon}},
\end{eqnarray}
suggesting a convenient representation of the posterior density matrix using $\hat{U}_{\epsilon}$,
\begin{eqnarray}
\hat{\rho}_{\epsilon}=\frac{1}{Z}\hat{U}_{\epsilon}\exp(\hat{\Lambda}_{\epsilon})\hat{U}_{\epsilon}^{\dag}.
\end{eqnarray}
In the limit $\epsilon\rightarrow 0$, the respective eigenvalues of $\hat{C}_{\epsilon}$, $\lambda_{\pm}$, approach $0$ and $-\infty$ while their respective eigenvectors straighten out $\ket{\lambda_{\pm}}_{\epsilon}\rightarrow\ket{\pm}$, and $\hat{U}_{\epsilon}\rightarrow\hat{1}$. Therefore the posterior density matrix $\hat{\rho}=\lim_{\epsilon\rightarrow 0}\hat{\rho}_{\epsilon}=\hat{\varphi}$ is equal to the biased prior density matrix and has not updated as $\hat{\varphi}_{\epsilon}\rightarrow \hat{\varphi}$. Because the pure state fails to update, it is biased, synonymous to a delta function probability distribution. The prior density matrix takes precedence as it does in the standard relative entropy case.  Below we will discuss the general case and discuss its implications.

Consider an $M$th order biased prior represented in its eigenbasis $\hat{\varphi}=\sum_{n=1}^M\varphi_n\ket{\varphi_n}\bra{\varphi_n}+\sum_{n=M+1}^N0_n\ket{\varphi_n}\bra{\varphi_n}$ in an $N$ dimensional Hilbert space ($M=1$ is a purestate). Given an $N\times N$ dimensional constraint $\hat{A}$ (however the analysis holds for $\hat{A}$ of any rank),  the prescription is to add and subtract some $\epsilon$'s from $\hat{\varphi}$ such that $\hat{\varphi}\rightarrow\hat{\varphi}_{\epsilon}$ spans $N$, and,
\begin{eqnarray} \log(\hat{\varphi}_{\epsilon})=\left( \begin{array}{cccccc}
\log(\varphi_1-\epsilon)& \dots & 0&0 &\dots&0\\
  \vdots&\ddots&0&0&\dots&0\\
0 & 0&\log(\varphi_M-\epsilon)&0&\dots&0\\
0 &0 &0 &\log(\epsilon)&&0\\
\vdots &\vdots & \vdots&&\ddots&0\\
0 &0 & 0&0&0&\log(\epsilon)
\end{array} \right),
\end{eqnarray}  
 has $N-M$ diagonal $\log(\epsilon)$ terms. Because density matricies are Hermitian, and have a sum representation, $\hat{\rho}=\sum_{ij}\rho_{ij}\ket{i}\bra{j}$, they may always be rearranged and relabeled into the form above without loss of generality. Note that this construction may not have $\mbox{Tr}(\varphi_{\epsilon})\neq \mbox{Tr}(\varphi)$, but there is no formal issue because equality holds in the limit $\epsilon\rightarrow 0$. Because $\log(\hat{\varphi}_{\epsilon})$ may always be reorganized as above, in general may be written as $\log(\hat{\varphi}_{\epsilon})=\log(\hat{\varphi}_M)\oplus\log(\epsilon)\hat{1}_{N-M}$, where $\log(\hat{\varphi}_M)$ is the first $M\times M$ block of $\log(\hat{\varphi}_{\epsilon})$ and $\log(\epsilon)\hat{1}_{N-M}$ is the remaining block proportional to $\log(\epsilon)$.  Expressing the $N\times N$ constraint matrices $\hat{A}=\sum_i\alpha_i\hat{A}_i$ in the eigenbasis of $\hat{\varphi}_{\epsilon}$, and summing it, is,
\begin{eqnarray}
\hat{C}_{\epsilon}=\hat{A}+\log(\hat{\varphi}_{\epsilon}),
\end{eqnarray}
which is a general representation of the matrix that residing in the exponential of a posterior density matrix, $\hat{\rho}_{\epsilon}=\frac{1}{Z}\exp(\hat{C}_{\epsilon})$, having an $M$th order biased prior density matrix $\hat{\varphi}_{\epsilon}$. If we similarly partition $\hat{C}_{\epsilon}$ by letting $\hat{C}_{M}$ be its first $M\times M$ block, then the characteristic polynomial equation of $\hat{C}_{\epsilon}$ has the following form,
 \begin{eqnarray}
0=\det|\hat{C}_{\epsilon}-\lambda\hat{1}|&=&\det|\hat{C}_{M}-\lambda\hat{1}_M|(\log(\epsilon)-\lambda)^{N-M}+c_{1}(\log(\epsilon)-\lambda)^{N-M-1}\nonumber\\
&+&c_{2}(\log(\epsilon)-\lambda)^{N-M-2}+...+c_{N-M},
\end{eqnarray}
where the $c_{n}$'s are the remaining coefficients of the characteristic polynomial. For any finite $\lambda$, we may divide the characteristic equation by the leading $(\log(\epsilon)-\lambda)^{N-M}\approx \log(\epsilon)^{N-M}$ term, which in the limit of $\epsilon\rightarrow 0$, reduces the characteristic equation to the $M\times M$ block characteristic equation, 
 \begin{eqnarray}
0=\det|\hat{C}_{\epsilon}-\lambda\hat{1}|\rightarrow\det|\hat{C}_{M}-\lambda\hat{1}_M|,
\end{eqnarray}
for all finite $\lambda$. The eigenvectors associated to these $M$-finite eigenvalues span the $M\times M$ vector space. As this is true for all finite eigenvalues, the remaining $N-M$ eigenvalues are not finite and indeed are all equal to negative infinity, due to the $\log(\epsilon)$'s as $\epsilon\rightarrow0$. The remaining eigenvectors with the associated infinite eigenvalues therefore span the remaining $(N-M)\times (N-M)$ vector space, but are not unique because they have degenerate eigenvalues. The eigenvectors for the finite and infinite eigenvalues are disjoint, and may be partitioned by a direct sum $\hat{\lambda}=\hat{\lambda}_M\oplus\hat{\lambda}_{N-M}$ and therefore so are the unitary matrices which diagonalize them $\hat{U}=\hat{U}_{M}\oplus\hat{U}_{N-M}$ as the unitary operators consist of columns of their associated eigenvectors. This disjointness is independence in the sense that the unitary operator $\hat{U}$ is block diagonal. The posterior density matrix is therefore,
\begin{eqnarray}
\hat{\rho}=\frac{1}{Z}e^{\hat{C}_M}\oplus\hat{0}_{N-M}=\frac{1}{Z}e^{\hat{A}_M+\log(\hat{\varphi}_M)}\label{biasedprior},
\end{eqnarray}
completely independent of the $\hat{A}-\hat{A}_M$ pieces of the constraints in $\hat{A}$, and $\hat{\varphi}_M=\hat{\varphi}$ is the original $M$th order biased prior. This means the expectation values used to constrain $\hat{\rho}$ should really be independent of the $\hat{A}-\hat{A}_M$ pieces to guarantee a logical solution. The lack of updating biased priors is not a failure of the method of maximum entropy, but rather a failure to choose appropriate constraints given an $M$th order biased prior density matrix -- essentially, this constraint and prior density matrix conflict and have no solution unless we change $\hat{A}\rightarrow\hat{A}_M$ such that $\mbox{Tr}(\hat{A}\hat{\rho})=\expt{A}\rightarrow\mbox{Tr}(\hat{A}_M\hat{\rho})=\expt{A_M}$. 

In general, any prior density matrix that does not span the entire Hilbert space is an $M$th order biased prior density matrix. This insists the following, which we state as a theorem:
\paragraph{Prior Density Matrix Theorem (PDMT):} An $M$th order biased prior density matrix $\hat{\varphi}$ can only be inferentially updated in the eigenspace that it spans. Regions not spanned by the $M$th order biased prior density matrix remain zero.\\

 The immediate consequence of the PDMT is that entropic updating can only cause \emph{epistemic and inferential} changes to $\hat{\rho}$. The inability to update a pure state, like in the pure state spin $\ket{+}\bra{+}$ example, shows just this. The only way to change the state of $\ket{+}\bra{+}$ is to physically rotate the state by applying dynamical unitary operators $U(t',t)$ via the Schr\"{o}dinger equation because no inferential entropic update is possible. Once the Quantum Bayes Rule (general collapse) is derived using entropy, we will see that the Schr\"{o}dinger equation and the quantum maximum entropy method compliment one another in QM -- the first being responsible for continuous dynamical ``physical" changes to the system and the second being discontinuous inferential updates, which cannot be explained by unitary evolution of the Schr\"{o}dinger equation. In some sense, asking questions like, ``What is the probability a spin up particle along $z$ is up along $x$?" is zero (at that time) unless it is further specified that something like a Stern-Gerlach device has been used to spatially separate (and decohere) the spin $\pm x$ values such they may be detected at a later time (in which case the answer is $\rho(\pm x|t')=1/2$). The Born Rule $\rho(x)=|\Psi(x)|^2$ seems to carry a lot of linguistic and experimental baggage if it is to be interpreted correctly. This is because detection, collapse, and entropic inference can only occur if first the pure state is projected into a mixed state $\hat{\rho}\rightarrow \sum \hat{P}_i\hat{\rho}\hat{P}_i$ by the appropriate measurement device. While it is possible for $\hat{\rho}$ and $\sum \hat{P}_i\hat{\rho}\hat{P}_i$ to have (in some basis) an identical probability spectrum, the two density matrices may evolve differently in time, and in that sense, represent different physical situations. 

If one is serious about the assignment of a biased prior density matrix then the following realization is needed, `` Because the prior density matrix is biased, the quantum maximum entropy method cannot update to a new posterior \emph{at that time}". If however your prior density matrix is changed physically by the addition of new microstates via interaction, allowing it to decohere \cite{Schlosshauer} (and the references therein), or change by some other process, then at a later time you one could employ a method similar to \cite{Wooters1,Wooters2}, that is, apply $\hat{\varphi}^{-1/2}\hat{\varphi}^{'1/2}(t)$ and its transpose on either side of $\hat{\varphi}\equiv\hat{\varphi}(0)$,
\begin{eqnarray}
\hat{\varphi}'(t)=\Big(\hat{\varphi}^{'1/2}(t)\hat{\varphi}^{-1/2}\Big)\,\hat{\varphi}\,\Big(\hat{\varphi}^{-1/2}\hat{\varphi}^{'1/2}(t)\Big),\label{changeofstate}
\end{eqnarray}
to represent a new prior density matrix that in general may: be a decohered, represent a new experimental configuration, or a unitary evolution. Now, if the prior is unbiased, it is possible to inferentially update it non-trivially. 

There are a few things to take away from this section. The quantum maximum entropy method only updates a density matrix inferentially, as can be seen by its lack of ability to rotate biased priors into non-biased states or other biased priors states. This is exactly what we expect, as the problem of biased priors exists in standard probability theory.   The solution to the biased prior problem is, if appropriate: to  change the constraint, change the prior, or perhaps both. This reasoning guides us in choosing appropriate priors in subsequent derivations throughout this paper. 

\section{The Quantum Bayes Rule:}
Notationally, we denote a density matrices living in a Hilbert space $\mathcal{H}_{x}\otimes\mathcal{H}_{\theta}$ to be written as $\hat{\rho}_{x,\theta}$. Density matrices may of course be expressed in any basis within these Hilbert spaces. We find it convenient to denote the $x',x'$ block matrix of $\hat{\rho}_{x,\theta}$ with an equal sign such that $\bra{x'}\hat{\rho}_{x,\theta}\ket{x'}\equiv\hat{\rho}_{x=x',\theta}$ and similarly $\bra{\theta'}\hat{\rho}_{x,\theta}\ket{\theta'}\equiv\hat{\rho}_{x,\theta=\theta'}$. Also a tilde above a density matrix will represent a mixed representation of the density matrix in question $\hat{\varphi}_{\theta}\rightarrow\stackrel{\sim}{\varphi}_{\theta}$. We introduce the Quantum Bayes Rule and then derive it using the quantum maximum entropy method as well as some generalizations.

\paragraph{Introduction -- Quantum Bayes Rule:}
Following \cite{Jacobs}, consider a prior density matrix $\hat{\varphi}_{\theta}$ which is entangled with an ancilla such that $\hat{\varphi}_{\theta}\rightarrow \hat{\varphi}_{x,\theta}$. The system and the ancilla are entangled in the following way; given an initial state of the ancilla $\ket{0}_x\bra{0}$, the joint system is entangled with a unitary operator $U$,
\begin{eqnarray}
\hat{\varphi}_{x,\theta}=U( \ket{0}_x\bra{0}\otimes\hat{\varphi}_{\theta})U^{\dag}
\end{eqnarray}
where,
\begin{eqnarray}
U=\sum_{\theta,\theta',x,x'}u_{\theta,\theta',x,x'}\ket{x'}\ket{\theta'}\bra{x}\bra{\theta}=\sum_{x,x'}\ket{x'}_x\bra{x}\otimes A_{x'x},
\end{eqnarray}
and,
\begin{eqnarray}
 A_{x'x}=\sum_{\theta,\theta'}u_{\theta,\theta',x,x'}\ket{\theta}\bra{\theta'},
\end{eqnarray}
is the $x',x$ sub-block matrix \cite{Jacobs}. The prior density matrix of the joint system is therefore,
\begin{eqnarray}
\hat{\varphi}_{x,\theta}=\sum_{x,x'}\ket{x}_x\bra{x'}\otimes (A_{x}\hat{\varphi}_{\theta}A^{\dag}_{x'})
\end{eqnarray}
where $A_{x0}\equiv A_{x}$ are defined as the measurement operators of the POVM $\hat{E}_{x}=\hat{A}_{x}\hat{A}^{\dag}_{x}$. Due to Neumark's theorem, making a projective measurement of the ancilla $x$ is a positive operator valued measurement (POVM) on $\hat{\varphi}_{\theta}$ \cite{Neumark}. Projecting the ancilla (is more or less collapse in the sense of L\"uders) requires the following action on $\hat{\varphi}_{x,\theta}$,
\begin{eqnarray}
\hat{\varphi}_{x,\theta}\rightarrow (\ket{x'}\bra{x'}\otimes I_{\theta})\hat{\varphi}_{x,\theta}(\ket{x'}\bra{x'}\otimes I_{\theta})=\ket{x'}\bra{x'}\otimes (A_{x'}\hat{\varphi}_{\theta}A^{\dag}_{x'}),
\end{eqnarray}
which implies the new state of the system is,
\begin{eqnarray}
\hat{\rho}_{\theta}=\frac{A_{x'}\hat{\varphi}_{\theta}A^{\dag}_{x'}}{\mbox{Tr}(A_{x'}\hat{\varphi}_{\theta}A^{\dag}_{x'})},\label{Qbayes2}
\end{eqnarray}
after normalizing, which is known as the Quantum Bayes Rule (QBR) \cite{Korotkov1,Korotkov2,Jordan}, the fundamental theorem of quantum measurement \cite{Jacobs}, or the POVM formalism \cite{Davies,Kraus,Holevo,Nielsen}. In the remainder of this section we will derive the Quantum Bayes Rule and other inference rules using the quantum maximum entropy method.

\paragraph{Simple Collapse:}
This entropic update is a special case of (\ref{Qbayes2}) when the $A_{x}$'s are all projectors rather than a more general POVM. As we are simply doing a projective measurement on $\hat{\varphi}_{x}$, an(other) ancilla is not needed to generate the POVM; however a projective measurement on the $x$'s requires entangling $\hat{\varphi}_{x}$ to detector states and letting them decohere within the detector. For concreteness we may imagine that $\hat{\varphi}_{x}$ represents the pure state density matrix of a particle that went though a two slit apparatus (no which slit measurement has been made) and is impeding onto a screen, CCD array, or the like. The pure state evolves with the detector states (as above),
\begin{eqnarray}
\ket{d_0}\bra{d_0}\otimes\hat{\varphi}_{x}\rightarrow\hat{\varphi}_{d,x}=\hat{U}\ket{d_0}\bra{d_0}\otimes\hat{\varphi}_{x}\hat{U}^{\dag},
\end{eqnarray}
and tracing over the detector states $\{\ket{d_i}\}$ to represent projective measurement on $x$ (before detection but after interaction with the measurement device), 
\begin{eqnarray}
\hat{\varphi}_x\rightarrow\stackrel{\sim}{\varphi}_{x}(t)=\mbox{Tr}_d(\hat{\varphi}_{d,x})=\sum_{x}\varphi(x)\ket{x}\bra{x},
\end{eqnarray}
is a mixed state realization of the original two slit pure state ($\varphi(x)=\stackrel{\sim}{\varphi}(x)$). This is in-no-way original and may be obtained following \cite{Luders} using projectors $\hat{\varphi}\rightarrow \stackrel{\sim}{\varphi}_{x}(t)=\sum_{x}\hat{P}_x\hat{\varphi}_x\hat{P}_x$ or more directly \cite{Schlosshauer}. 

In principle, when the detection of the result of a projective measurement ($\stackrel{\sim}{\varphi}_{x}$) is made, the state of the system is known with certainty. This is represented by the following constraint on the posterior probability distribution, 
\begin{eqnarray}
\mbox{Tr}(\ket{x}\bra{x}\hat{\rho}_{x})=\rho(x)=\delta_{xx'},\label{delta}
\end{eqnarray}
which is an expectation value on the posterior density matrix $\hat{\rho}_{x}$, stating that the system was detected in the $x'$ state with certainty. Because this constraint must be imposed for every $x$, there is one Lagrange multiplier $\alpha_x$ for each $x$. Maximizing the quantum relative entropy with respect to this constraint and normalization is setting
\begin{eqnarray}
0=\delta\Big( S-\lambda[\mbox{Tr}(\hat{\rho}_{x})-1]-\sum_x\alpha_x[\mbox{Tr}(\ket{x}\bra{x}\hat{\rho}_{x})-\delta_{xx'}]\Big),\label{47}
\end{eqnarray}
which gives the posterior,
 \begin{eqnarray}
\hat{\rho}_{x}=\frac{1}{Z}\exp\Big(\sum_x\alpha_x\ket{x}\bra{x}+\log(\stackrel{\sim}{\varphi}_{x})\Big).
\end{eqnarray}
Because the constraint and prior commute, the posterior density matrix takes a simple form,
\begin{eqnarray}
\hat{\rho}_{x}=\frac{1}{Z}\sum_x\varphi(x)e^{\alpha_x}\ket{x}\bra{x}.
\end{eqnarray}
The normalization constraint gives,
\begin{eqnarray}
Z=\mbox{Tr}(\hat{\rho}_{x})=\sum_x\varphi(x)e^{\alpha_x},
\end{eqnarray}
and the expectation value constraint (\ref{delta}) gives,
\begin{eqnarray}
\delta_{xx'}=\mbox{Tr}(\ket{x}\bra{x}\hat{\rho}_{x})=\frac{\varphi(x)e^{\alpha_x}}{Z}.
\end{eqnarray}
The final form of the posterior density matrix is found by substituting for $\alpha_x$, and the result is a collapsed state is,
\begin{eqnarray}
\hat{\rho}_{x}=\sum_x\delta_{xx'}\ket{x}\bra{x}=\ket{x'}\bra{x'},
\end{eqnarray}
as suspected.  Written in a suggestive ``Bayes update" or "projective collapse" form,
\begin{eqnarray}
\hat{\rho}_{x}\equiv\,\,\stackrel{\sim}{\varphi}_{x|x'}\,\,=\frac{\ket{x'}\bra{x'}\stackrel{\sim}{\varphi}_{x}\ket{x'}\bra{x'}}{\varphi(x')}
=\frac{\ket{x'}\stackrel{\sim}{\varphi}_{x=x'}\bra{x'}}{\varphi(x')},
\end{eqnarray}
it perhaps better meshes L\"{u}ders strong collapse rule and the QBR. Note the tilde on $\stackrel{\sim}{\varphi}_{x}$ indicates that it is the appropriate prior for inference as the state has decohered in the detector. Although $\stackrel{\sim}{\varphi}_{x=x'}$ and $\hat{\varphi}_{x=x'}$ are numerically equal, substitution of this above is incorrect because $\hat{\varphi}_{x}$ has yet to decohere and cannot be inferentially updated due to the PDMT. Although this is perhaps a bit fussy, it provides another reason why secure channels exist in quantum cryptography -- the statistics and dynamics of a quantum system change when it is measured ($\hat{\varphi}_x\rightarrow\stackrel{\sim}{\varphi}_{x}$) because the state must decohere before it is inferentially updated ($\stackrel{\sim}{\varphi}_{x}\rightarrow \hat{\rho}_{x}$) due to the PDMT. 

Above is the special case of the Quantum Bayes Rule (\ref{Qbayes}) when the measurements are projective. Note that this derivation does not require first solving for the ``weak" collapse and taking the limit as is done in \cite{Hellmann} to avoid infinite relative entropies \cite{Hellmann}. This is because \cite{QRE} gives the general solution to $\hat{\rho}$ (equation (\ref{rhosolution})) while also providing the quantum maximum entropy method for making inferential updates of density matrices.   

\paragraph{Simple Weak Collapse:}
A form of weak collapse may be found by considering a system that has a certain probability of being in one state or another (perhaps due to measurement uncertainty) after detection. Given the same prior density matrix $\stackrel{\sim}{\varphi}_{x}$, we maximize the entropy with respect to a set of constraints $\{\rho(x)=\mbox{Tr}(\ket{x}\bra{x}\hat{\rho}_{x})\}$ to codifying a lack of certainty in the state (perhaps a narrow Gaussian distribution rather than exact knowledge in (\ref{delta})). Maximizing the entropy with respect to these constraints and normalization again gives the posterior,
\begin{eqnarray}
\hat{\rho}_{x}=\frac{1}{Z}\sum_{x}\varphi(x)e^{\alpha_{x}}\ket{x}\bra{x},
\end{eqnarray}
because all the matrices and projectors commute. The normalization constraint gives,
\begin{eqnarray}
Z=\mbox{Tr}(\hat{\rho}_{x})=\sum_{x}\varphi(x)e^{\alpha_{x}}.
\end{eqnarray}
Satisfying the remaining expectation value constraint ($\rho(x)=\mbox{Tr}(\ket{x}\bra{x}\hat{\rho}_{x})$) gives $e^{\alpha_{x}}=Z\frac{\rho(x)}{\varphi(x)}$ for each $x$, and therefore,
\begin{eqnarray}
\hat{\rho}_{x}=\sum_{x}\rho(x)\ket{x}\bra{x}=\sum_{x'}\rho(x')\frac{\ket{x'}\stackrel{\sim}{\varphi}_{x=x'}\bra{x'}}{\varphi(x')}=\sum_{x'}\rho(x')\stackrel{\sim}{\varphi}_{x|x'},
\end{eqnarray}
which is a weak collapse or perhaps a quantum Jeffrey's rule in agreement with \cite{Hellmann}. 

\paragraph{The Appropriate Prior for the QBR:}
The problem at hand requires a knowledge of the correct prior density matrix for inference. 
%
%
 Notice that if $\hat{\varphi}_{\theta}$ is an $M$th order biased prior, then,
\begin{eqnarray}
\hat{\varphi}_{x,\theta}=U( \ket{0}_x\bra{0}\otimes\hat{\varphi}_{\theta})U^{\dag},
\end{eqnarray}
is an $M$th order biased prior, meaning that $\hat{\varphi}_{x,\theta}$ can only be inferentially updated in that subspace (which may or may not be desirable). This is especially problematic if $M=1$ such that $\hat{\varphi}_{x,\theta}=\hat{\varphi}_{x,\theta}^2$ is a pure state because it cannot be updated at all. 

We therefore follow the intuition given by the PDMT -- if we are going to make inferences on the basis of detection, the prior density matrix should appropriately reflect the fact that it has interacted with a measurement device. This interaction will be modeled by entangling the ancilla and detector states $\{\ket{d_y}\}$, which act as a local environment states within the detector, via a unitary evolution (following \cite{Schlosshauer} and the notation in \cite{Jacobs}, but a simple projection argument from L\"uders on the ancilla states of $\varphi_{x\theta}$ would also suffice), 
\begin{eqnarray}
\ket{d_0}\bra{d_0}\otimes\hat{\varphi}_{x,\theta}\rightarrow \hat{\varphi}_{d,x,\theta}=(\hat{U}_{dx}\otimes I_{\theta})(\ket{d_0}\bra{d_0}\otimes\hat{\varphi}_{x,\theta})(\hat{U}^{\dag}_{dx}\otimes I_{\theta})
\end{eqnarray}
where,
\begin{eqnarray}
\hat{U}_{dx}=\sum_{d_y,d_{y'},x,x'}u_{d_yd_{y}xx'}\ket{d_y}\ket{x}\bra{d_{y'}}\bra{x'}=\sum_{d_y,d_{y'}}\ket{d_y}\bra{d_{y'}}\otimes B_{d_yd_{y'}},
\end{eqnarray}
and the sub-block matrices are,
\begin{eqnarray}
B_{d_yd_{y'}}=\sum_{x,x'}u_{d_yd_{y'}xx'}\ket{x}\bra{x'}.
\end{eqnarray}
We define a good detector as one in which the $\ket{x}$th ancilla state only entangles with the local state of the detector $\ket{d_x}$, which is an argument for the sub-block matrix to take a simple form,
\begin{eqnarray}
B_{d_y0}=\sum_{x,x'}\delta_{y,x}\delta_{y,x'}\ket{x}\bra{x'}=\ket{y}\bra{y}.
\end{eqnarray}
The entangled density matrix becomes,
\begin{eqnarray}
\hat{\varphi}_{d,x,\theta}=\sum_{y,y',x,x'}\ket{d_y}\bra{d_{y'}}\otimes(B_{y'0}\otimes I_{\theta})\hat{\varphi}_{x,\theta}(B_{y'0}^{\dag}\otimes I_{\theta})\nonumber\\
=\sum_{y,y'}\ket{d_y}\bra{d_{y'}}\otimes\ket{y}\bra{y'}\otimes  (A_{y}\stackrel{\sim}{\varphi}_{\theta}A^{\dag}_{y'}).
\end{eqnarray}
These local environment detector states in which the ancilla reside, are traced over, as we do not keep track of their evolution. An example of this would be an ancilla which terminates on a photosensitive sheet -- we obviously do not keep track of the state of the sheet.  This is to say, a small period of time after the projective measurement has been made, the ancilla states transitions to a mixed state, which gives a standard (classical) probability distribution of the ancilla states over the detector. The prior density matrix after projective measurement has been made is thus a block diagonal sum of states,
\begin{eqnarray}
\stackrel{\sim}{\varphi}_{x,\theta}(t)=\mbox{Tr}_{d}(\hat{\varphi}_{d,x,\theta})=\sum_x\ket{x}\bra{x}\otimes  (A_{x}\hat{\varphi}_{\theta}A^{\dag}_{x})=\sum_{x'}\ket{x'}\bra{x'}\otimes  \stackrel{\sim}{\varphi}_{x=x',\theta},
\end{eqnarray}
which we claim is the appropriate density matrix for POVM inference. This form of the prior is nolonger biased, even if $\hat{\varphi}_{\theta}$ is itself biased. If all of the $A_{x}$'s are projections, then this prior represents the resulting mixed state from a detector interacting with a potentially entangled of a pure state such as $\ket{\Psi}=\sum_x c_x\ket{x,\theta_x}$. As is done in \cite{Luders}, the action of the measurement device causes $\hat{\varphi}_{x,\theta}\rightarrow\stackrel{\sim}{\varphi}_{x,\theta}$, but here this change of state is required to make inferential updates due to the PDMT, complimenting L\"{u}ders.

\paragraph{The constraints leading to QBR:}
Detecting the (exact) result of a projective measurement on the ancilla state $x$ puts the posterior ancilla into a collapsed state $x'$. This is represented by a posterior probability distribution (data) expectation value constraint, 
\begin{eqnarray}
\rho(x)=\mbox{Tr}(\ket{x}\bra{x}\otimes\hat{1}_{\theta}\cdot\hat{\rho}_{x,\theta})=\delta_{xx'},\label{delta2}
\end{eqnarray}
for the case when the final state of the ancilla is known. Notice that this information alone is not enough to fully constrain $\hat{\rho}_{x,\theta}$ as there are many $\hat{\rho}_{x,\theta}$ which satisfy that constraint. We therefore employ the quantum maximum entropy method and impose normalization, this data constraint, with respect to the appropriate prior $\stackrel{\sim}{\varphi}_{x,\theta}$,
\begin{eqnarray}
0=\delta\Big( S-\lambda[\mbox{Tr}(\hat{\rho}_{x,\theta})-1]-\sum_x\alpha_x[\mbox{Tr}(\ket{x}\bra{x}\otimes\hat{1}_{\theta}\cdot\hat{\rho}_{x,\theta})-\delta_{xx'}]\Big),
\end{eqnarray}
which gives,
 \begin{eqnarray}
\hat{\rho}_{x,\theta}=\frac{1}{Z}\exp\Big(\sum_x\alpha_x\ket{x}\bra{x}\otimes\hat{1}_{\theta}+\log(\stackrel{\sim}{\varphi}_{x,\theta})\Big).
\end{eqnarray}
Because the prior density matrix is block diagonal $\log(\stackrel{\sim}{\varphi}_{x,\theta})=\sum_x\ket{x}\bra{x}\otimes\log(A_{x}\hat{\varphi}_{\theta}A^{\dag}_{x})=\sum_x\ket{x}\bra{x}\otimes\log(\stackrel{\sim}{\varphi}_{x=x,\theta})$ we have,
\begin{eqnarray}
\hat{\rho}_{x,\theta}=\frac{1}{Z}\sum_x\ket{x}\bra{x}\otimes e^{\alpha_x\hat{1}_{\theta}+\log(\stackrel{\sim}{\varphi}_{x=x,\theta})}=\frac{1}{Z}\sum_xe^{\alpha_x}\ket{x}\bra{x}\otimes\stackrel{\sim}{\varphi}_{x=x,\theta}.
\end{eqnarray}
Imposing normalization gives,
\begin{eqnarray}
Z=\mbox{Tr}\Big(\sum_xe^{\alpha_x}\ket{x}\bra{x}\otimes\stackrel{\sim}{\varphi}_{x=x,\theta}\Big)=\sum_xe^{\alpha_x}\sum_{\theta}\bra{\theta}\stackrel{\sim}{\varphi}_{x=x,\theta}\ket{\theta}=\sum_xe^{\alpha_x}\stackrel{\sim}{\varphi}(x).
\end{eqnarray}
The expectation value constraint forces,
\begin{eqnarray}
\rho(x)=\frac{1}{Z}\mbox{Tr}(\ket{x}\bra{x}\otimes\hat{1}_{\theta}\,\,\sum_x e^{\alpha_x}\ket{x}\bra{x}\otimes\stackrel{\sim}{\varphi}_{x=x,\theta})\nonumber\\
=\frac{1}{Z}\mbox{Tr}(e^{\alpha_x}\ket{x}\bra{x}\otimes\stackrel{\sim}{\varphi}_{x=x,\theta})=\frac{e^{\alpha_x}}{Z}\sum_{\theta}\bra{\theta}\stackrel{\sim}{\varphi}_{x=x,\theta}\ket{\theta}=\frac{e^{\alpha_x}}{Z}\stackrel{\sim}{\varphi}(x),
\end{eqnarray}
meaning, $e^{\alpha_x}=\frac{Z\rho(x)}{\stackrel{\sim}{\varphi}(x)}$. In the case the data is known exactly, $\rho(x)=\delta_{xx'}$, the Lagrange multiplier reads $e^{\alpha_x}=\frac{Z\delta_{xx'}}{\stackrel{\sim}{\varphi}(x)}$. Substituting in for the Lagrange multipliers gives the final form of the posterior density matrix ,
\begin{eqnarray}
\hat{\rho}_{x,\theta}=\sum_x\frac{\delta_{xx'}}{\stackrel{\sim}{\varphi}(x)}\ket{x}\bra{x}\otimes\stackrel{\sim}{\varphi}_{x=x,\theta}=\frac{\ket{x'}\bra{x'}\otimes\stackrel{\sim}{\varphi}_{x=x',\theta}}{\stackrel{\sim}{\varphi}(x')}.
\end{eqnarray}
The marginal posterior gives the Quantum Bayes Rule,
\begin{eqnarray}
\hat{\rho}_{\theta}=\mbox{Tr}_{x}(\hat{\rho}_{x,\theta})=\frac{\stackrel{\sim}{\varphi}_{x=x',\theta}}{\stackrel{\sim}{\varphi}(x')}\equiv \stackrel{\sim}{\varphi}_{\theta|x'},
\end{eqnarray}
which is equivalent to the standard measurement operator $\hat{A}_{x}$ formulation,
\begin{eqnarray}
\hat{\rho}_{\theta}=\frac{A_{x'}\hat{\varphi}_{\theta}A^{\dag}_{x'}}{\varphi(x')},
\end{eqnarray}
of the QBR. The posterior probability of $\theta$ indeed gives the standard Bayes Rule,
  \begin{eqnarray}
\rho(\theta)=\mbox{Tr}(\hat{\rho}_{\theta}\ket{\theta}\bra{\theta})=\mbox{Tr}(\stackrel{\sim}{\varphi}_{\theta|x'}\ket{\theta}\bra{\theta})=\varphi(\theta|x')=\frac{\varphi(\theta,x')}{\varphi(x')},
\end{eqnarray}
as density matrices generate probability distributions in this fashion. As stated in \cite{Jordan}, the off diagonal elements in $\hat{\rho}_{\theta}$ have a more exotic updating rule.\\
\paragraph{Quantum Jeffrey's Rule (QJR):} In the same way as before, we may easily generalize this rule to cases in which the final state of the ancilla is uncertain and encoded by a probability distribution $\rho(x)$ rather than one exhibiting certainty $\delta_{xx'}$. Simply replacing the expectation value constraint (\ref{delta2}) by,
\begin{eqnarray}
\mbox{Tr}(\ket{x}\bra{x}\otimes\hat{1}_{\theta}\cdot\hat{\rho}_{x,\theta})=\rho(x),
\end{eqnarray}
and performing the quantum maximum entropy method gives the marginal posterior,
\begin{eqnarray}
\hat{\rho}_{\theta}=\mbox{Tr}_{x}(\hat{\rho}_{x,\theta})=\sum_x\rho(x) \stackrel{\sim}{\varphi}_{\theta|x}=\sum_x\rho(x)\frac{A_{x}\hat{\varphi}_{\theta}A^{\dag}_{x}}{\varphi(x)},
\end{eqnarray}
which gives a Quantum Jeffrey's Rule for POVMs.

\paragraph{What about $ \stackrel{\sim}{\vartheta}_{x|\theta'}$? } 

Now that we have a decent definition of $\stackrel{\sim}{\varphi}_{\theta|x'}$, it is reasonable to inquire if a complement density matrix $\stackrel{\sim}{\vartheta}_{x|\theta'}$ can be defined for the same system and interaction scheme.

Consider rewriting the entangled ancilla prior density matrix $\hat{\varphi}_{x,\theta}$ as,
\begin{eqnarray}
\hat{\varphi}_{x,\theta}=\sum_{x,x'}\ket{x}_x\bra{x'}\otimes (A_{x}\hat{\varphi}_{\theta}A^{\dag}_{x'})=\sum_{\theta,\theta'}(A_{\theta}\hat{\vartheta}_xA^{\dag}_{\theta'})\otimes\ket{\theta}_{\theta}\bra{\theta'},
\end{eqnarray}
where the $A_{\theta}$'s are a bit messy but obtained from moving and relabeling the components of the unitary matrices.

We have done this such that we may make a projective measurement on $\theta$ and infer the ancilla state. This involves decohering the $\theta$ states locally inside a $\theta$ projection measurement device having states $\{\ket{d_{\theta}}\}$. Making analogous arguments from before, we trace over the entangled detector states and obtain the appropriate prior density matrix for inference,
 \begin{eqnarray}
\hat{\varphi}_{x,\theta}\rightarrow\stackrel{\sim}{\vartheta}_{x,\theta}=\sum_{\theta}(A_{\theta}\hat{\vartheta}_xA^{\dag}_{\theta})\otimes\ket{\theta}_{\theta}\bra{\theta},
\end{eqnarray}
which is block diagonal in $\ket{\theta}$. It should be noted that in general $\stackrel{\sim}{\vartheta}_{x,\theta}\neq\stackrel{\sim}{\varphi}_{x,\theta}$, from the previous section, as they are block diagonal in different Hilbert spaces. The same analysis from the previous section is made: using $\stackrel{\sim}{\vartheta}_{x,\theta}$ as the prior, maximize the entropy with respect to normalization, the data constraint
\begin{eqnarray}
\mbox{Tr}(\hat{1}_x\otimes\ket{\theta}\bra{\theta}\cdot\hat{\varrho}_{x,\theta})=\delta_{\theta\theta'},
\end{eqnarray}
and solve for the Lagrange multipliers. This gives
 \begin{eqnarray}
\hat{\varrho}_{x,\theta}=\frac{\stackrel{\sim}{\vartheta}_{x,\theta=\theta'}}{\vartheta(\theta')}\otimes\ket{\theta'}\bra{\theta'},
\end{eqnarray}
such that the marginal posterior is
\begin{eqnarray}
\hat{\varrho}_{x}\equiv \stackrel{\sim}{\vartheta}_{x|\theta'}=\mbox{Tr}_{\theta}(\hat{\varrho}_{x,\theta})=\frac{\stackrel{\sim}{\vartheta}_{x,\theta=\theta'}}{\vartheta(\theta')}\,,
\end{eqnarray}
which is equivalent to,
\begin{eqnarray}
\hat{\varrho}_{x}=\frac{A_{\theta'}\hat{\vartheta}_{x}A^{\dag}_{\theta'}}{\vartheta(\theta')},
\end{eqnarray}
and is interpreted as the posterior density matrix of the ancilla after a complementary $\hat{A}_{\theta'}$ measurement operator has been applied and $\theta'$ has been detected.

 The conditional priors, $\stackrel{\sim}{\vartheta}_{x|\theta'}$ and $ \stackrel{\sim}{\varphi}_{\theta|x'}$, are related in the following way: Notice that although $\stackrel{\sim}{\vartheta}_{x,\theta}\neq\stackrel{\sim}{\varphi}_{x,\theta}$ in general, their components along the diagonal-diagonal are equal, $\bra{x,\theta}\stackrel{\sim}{\vartheta}_{x,\theta}\ket{x,\theta}=\bra{x,\theta}\stackrel{\sim}{\varphi}_{x,\theta}\ket{x,\theta}=\bra{x,\theta}\hat{\varphi}_{x,\theta}\ket{x,\theta}$, because both density matrices are decohered in one way or another from the same original entangled prior density matrix $\hat{\varphi}_{x,\theta}$. This gives the following relationship among their probability distributions,
\begin{eqnarray}
\varphi(x,\theta)=\vartheta(x,\theta),
\end{eqnarray}
and because $\varphi(x)=\vartheta(x)$ and $\varphi(\theta)=\vartheta(\theta)$,
\begin{eqnarray}
\varphi(\theta|x)=\frac{\vartheta(x,\theta)}{\varphi(x)}=\frac{\vartheta(x,\theta)}{\vartheta(x)}=\frac{\vartheta(x|\theta)\vartheta(\theta)}{\vartheta(x)}=\vartheta(\theta|x),
\end{eqnarray}
and likewise $\vartheta(x|\theta)=\varphi(x|\theta)$, we see all of the probability relationships hold and may be used interchangeably. It should also be noted that in general: $\stackrel{\sim}{\vartheta}_{x}\neq\stackrel{\sim}{\varphi}_{x}$ and $\stackrel{\sim}{\vartheta}_{\theta}\neq\stackrel{\sim}{\varphi}_{\theta}$ because their off diagonal components may differ, however you may express $\stackrel{\sim}{\vartheta}_{x|\theta'}$ in terms of the $\varphi$ probability distributions and vice-versa. The joint posterior density matricies $\stackrel{\sim}{\vartheta}_{x,\theta}$ and $\stackrel{\sim}{\varphi}_{x,\theta}$, and their posterior marginals differ in how they will evolve in time. It is also possible to make inferences on a prior state in which both Hilbert spaces have decohered, $\hat{\varphi}_{x\theta}\rightarrow \,\sum_{x,\theta}\varphi(x,\theta)\ket{x,\theta}\bra{x,\theta}$, which has correlations due to the previous entanglement but is no longer entangled. Because the use of appropriate measurement devices leads to $\varphi(x,\theta)=\vartheta(x,\theta)$, there is no interpretational issue in the delayed choice experiment because collapse only occurs after detection and decoherence of both the ancilla $x$ and system of interest $\theta$. The time order of the decoherence becomes irrelevant because the joint probabilities are equal -- a similar argument is given in \cite{Gaasbeek}. Essentially what has happened in the delayed choice experiment is that you do not know if you have done a ``which slit" measurement or not, which is like having ``mixed state of measurement outcomes", but, this is precisely what a POVM measurement represents.


\paragraph{Weak collapse via thermal baths:}
Rather than detecting the result of a projective measurement on the ancilla state, we consider the weak measurement POVM one would obtain if the ancilla is sent into a thermal box as it can be naturally generated in the quantum maximum entropy method. Here we will let the Hilbert space $\mathcal{H}_x$ of the ancilla be spanned by $\{\ket{n}\}$, the energy basis eigenstates of the ancilla in the thermal box having a Hamiltonian $\hat{H}_{n}=\sum_n\epsilon_n\ket{n}\bra{n}$. The joint prior density matrix is prepared similar to above $\stackrel{\sim}{\varphi}_{x,\theta}\equiv \stackrel{\sim}{\varphi}_{n,\theta}$ such that,
\begin{eqnarray}
\stackrel{\sim}{\varphi}_{n,\theta}=\sum_{n'}\ket{n'}\bra{n'}\otimes  \stackrel{\sim}{\varphi}_{n=n',\theta}.
\end{eqnarray}
 The following energy expectation value is used to represent the constraint of an ancilla in a thermal box,
\begin{eqnarray}
\mbox{Tr}(\hat{H}_{n}\otimes\hat{1}_{\theta}\cdot\hat{\rho}_{n,\theta})=\expt{H_{n}}.
\end{eqnarray}
Again notice that this information alone is not enough to fully constrain $\hat{\rho}_{n,\theta}$ as there are many $\hat{\rho}_{n,\theta}$ which satisfy that constraint. We therefore require the quantum maximum entropy method; that is, maximizing the quantum relative entropy with respect to normalization, this constraint, and the POVM prior $\stackrel{\sim}{\varphi}_{n,\theta}$ is,
\begin{eqnarray}
0=\delta\Big( S-\lambda[\mbox{Tr}(\hat{\rho}_{n,\theta})-1]-\beta[\mbox{Tr}(\hat{H}_{n}\otimes\hat{1}_{\theta}\cdot\hat{\rho}_{n,\theta})-\expt{H_n}]\Big),
\end{eqnarray}
which gives,
 \begin{eqnarray}
\hat{\rho}_{n,\theta}=\frac{1}{Z}\exp\Big(\beta\hat{H}_{n}\otimes\hat{1}_{\theta}+\log(\stackrel{\sim}{\varphi}_{n,\theta})\Big).
\end{eqnarray}
Because the POVM prior density matrix is block diagonal $\log(\stackrel{\sim}{\varphi}_{n,\theta})=\sum_{n}\ket{n}\bra{n}\otimes\log(A_{n}\hat{\varphi}_{\theta}A^{\dag}_{n})=\sum_{n}\ket{n}\bra{n}\otimes\log(\stackrel{\sim}{\varphi}_{n=n,\theta})$ we have that,
\begin{eqnarray}
\hat{\rho}_{n,\theta}=\frac{1}{Z}\sum_{n}\ket{n}\bra{n}\otimes e^{\beta\epsilon_n\hat{1}_{\theta}+\log(\stackrel{\sim}{\varphi}_{n=n,\theta})}=\frac{1}{Z}\sum_{n}e^{\beta \epsilon_n}\ket{n}\bra{n}\otimes\stackrel{\sim}{\varphi}_{n=n,\theta}.
\end{eqnarray}
Imposing normalization gives,
\begin{eqnarray}
Z=\mbox{Tr}\Big(\sum_{n}e^{\beta \epsilon_n}\ket{n}\bra{n}\otimes\stackrel{\sim}{\varphi}_{n=n,\theta}\Big)=\sum_n e^{\beta \epsilon_n}\sum_{\theta}\bra{\theta}\stackrel{\sim}{\varphi}_{n=n,\theta}\ket{\theta}=\sum_{n} e^{\beta \epsilon_n}\stackrel{\sim}{\varphi}(\epsilon_n).
\end{eqnarray}
The expectation value constraint forces,
\begin{eqnarray}
\expt{H_{n}}=\mbox{Tr}(\hat{H}_{n}\otimes\hat{1}_{\theta}\cdot\hat{\rho}_{n,\theta})=\frac{1}{Z}\mbox{Tr}(\hat{H}_{n}\otimes\hat{1}_{\theta}\,\,\sum_{n}e^{\beta \epsilon_n}\ket{n}\bra{n}\otimes\stackrel{\sim}{\varphi}_{n=n,\theta})\nonumber\\
=\frac{1}{Z}\mbox{Tr}(\sum_{n}\epsilon_n e^{\beta \epsilon_n}\ket{n}\bra{n}\otimes\stackrel{\sim}{\varphi}_{n=n,\theta})=\sum_{n}\frac{\epsilon_n e^{\beta \epsilon_n}}{Z}\stackrel{\sim}{\varphi}(\epsilon_n)=\frac{\partial }{\partial \beta}\log (Z).
\end{eqnarray}
meaning one can solve $\beta=\beta(\expt{H_{n}})$ by inverting the above equation after computing $Z$ as is done in Statistical Mechanics. The marginal posterior is a realization of the ``weak" collapse rule using thermalization,
\begin{eqnarray}
\hat{\rho}_{\theta}=\mbox{Tr}_{n}(\hat{\rho}_{n,\theta})=\sum_{n}\frac{e^{\beta \epsilon_n}}{Z}\stackrel{\sim}{\varphi}_{n=n,\theta}=\sum_{n}\frac{\varphi(\epsilon_n)e^{\beta \epsilon_n}}{Z}\stackrel{\sim}{\varphi}_{\theta|n}=\sum_{n}\rho(\epsilon_n)\stackrel{\sim}{\varphi}_{\theta|n}=\sum_{n}\rho(\epsilon_n)\frac{A_{n}\stackrel{\sim}{\varphi}_{\theta}A^{\dag}_{n}}{\varphi(\epsilon_n)},
\end{eqnarray}
in which the outcome state $\hat{\rho}_{\theta}$ of the system may be controlled (in the usual sense) by forcing the ancilla into a box with temperature $\beta$ or $\beta'$ as it causes changes to the statistics of the distant weak POVM: $\rho(\theta)=\mbox{Tr}(\ket{\theta}\bra{\theta}\hat{\rho}_{\theta})$. 

\subsection{Generalizations:}
General inferences of $\hat{\rho}$ on the basis of a prior state of knowledge $\hat{\varphi}$ and arbitrary expectation value constraints $\{\expt{\hat{A}_i}\}$ gives the following general updating rule,
\begin{eqnarray}
\hat{\rho}=\frac{1}{Z}\exp\Big(\sum_i\alpha_i\hat{A}_i+\ln(\hat{\varphi})\Big),
\end{eqnarray}
from $\hat{\varphi}\rightarrow \hat{\rho}$ in light of new information about $\{\expt{\hat{A}_i}\}$. This is of-course the general solution to the quantum maximum entropy method, but now it is clear it may be interpreted as the solution for general purpose inference when applied correctly. As commutation was used in the previous QBR and QJR, inferences involving expectation values of non-commuting operators generalizes these rules -- for instance, ``simultaneously" imposing $\expt{\hat{x}}$ and $\expt{\hat{p}}$ gives $\hat{\rho}=\frac{1}{Z}\exp\Big(\alpha\hat{x}+\beta\hat{p}+\log(\hat{\varphi})\Big)$. The solution is found by diagonalizing the exponential and using the methods from Statistical Mechanics.

An odd prescription might be, given a prior density matrix which has decohered or is being measured in the of $\hat{A}$ basis (due to the PDMT) such that,
\begin{eqnarray}
\hat{\varphi}\rightarrow\, \stackrel{\sim}{\varphi}=\sum_a\ket{a}\hat{\varphi}_{a=a}\bra{a}=\sum_a\ket{a}\varphi(a)\bra{a},
\end{eqnarray} 
consider maximizing the entropy with respect to the expectation value of an operator $\hat{B}$,
\begin{eqnarray}
\expt{\hat{B}}=\mbox{Tr}(\hat{B}\cdot\hat{\rho}),
\end{eqnarray}
that does not commute $[\hat{A},\hat{B}]\neq0$ (e.g. considering momentum expectation value constraints on a system projected onto a position measurement device). The solution given by the quantum maximum entropy method is,
\begin{eqnarray}
\hat{\rho}=\frac{1}{Z}\exp\Big(\alpha\hat{B}+\ln(\stackrel{\sim}{\varphi})\Big)\equiv\frac{1}{Z}\exp(\hat{C}).
\end{eqnarray}
To find $Z$, diagonalize the Hermitian matrix $\hat{C}\rightarrow \hat{\Lambda}=\sum_{\lambda}\lambda\ket{\lambda}\bra{\lambda}$, such that,
\begin{eqnarray}
Z=\mbox{Tr}(\exp(\hat{C}))=\mbox{Tr}(U\exp(U^{\dag}\hat{C}U)U^{\dag})=\mbox{Tr}(\exp(\hat{\Lambda}))=\sum_{\lambda} e^{\lambda}.
\end{eqnarray}
 Finding $\alpha$ is then reduced to methods from Statistical Mechanics,
\begin{eqnarray}
\expt{\hat{B}}=\mbox{Tr}(\hat{B}\cdot\hat{\rho})=\frac{\partial}{\partial \alpha}\log (Z),
\end{eqnarray}
which may be inverted to find $\alpha=\alpha(\expt{\hat{B}})$, and thus the posterior $\hat{\rho}$. 

Perhaps the simplest example of such a situation is to start from a mixed prior density matrix in spin-z and then maximizing the quantum relative entropy with respect to expectations in spin-x to infer the posterior density matrix. The quantum maximum entropy method reproduces the well known solution to this problem that is usually reasoned by appropriately weighting the eigenvalues of $\hat{\rho}$ (in the $x$ basis) such that $\expt{\sigma_x}$ is satisfied (the solution is completely determined due to normalization). The quantum maximum entropy method may be extended to cases in which the system of equations is under-determined, e.g. a single expectation value constraint and normalization of a Gell-Mann operator constraint $\expt{\hat{\lambda}_i}$ that does not commute with its $3\times 3$ (mixed) prior density matrix.

\section{Conclusions:}
In this article we applied the quantum maximum entropy method and derived the L\"{u}ders collapse (and weak collapse) rules, the QBR, the QJR, and also a method for computing inferential generalizations when expectation values do not commute.
In doing so we eliminated \emph{ad hoc} collapse postulates in QM by using the quantum maximum entropy method \cite{QRE}. As is demonstrated by the arguments leading up to the PDMT, because $M$th order biased priors may only be inferentially updated (full or partial collapse) within the $M$ dimensional Hilbert subspace they span, the phrase, ``collapse of the wavefunction" should be replaced by ``collapse of the mixed state". A simple consequence is, because an $M=1$ biased prior (pure state) cannot be updated inferentially, as given by the PDMT, it shows that entropic methods provide a reason for why pure states are ``secure channels" simply because any eavesdropper would have to decohere the pure state to make inferences, and change the statistics of the original state.  In essence, the PDMT is a rediscovery of L\"{u}ders notion that the application of a measurement device is to mix the incoming state $\hat{\rho}\rightarrow \sum_{i}\hat{P}_i\hat{\rho}\hat{P}_i$, except here it was derived purely from inferential and entropic arguments. The quantum maximum entropy method and the PDMT have rigorized some notions and applications of quantum measurement such that future applications have a more full bodied representation in Quantum Theory.

\section{Acknowledgments}
We would to like the reviewers and also: N. Carrara, T. Gai, S. Ipek, J. Tian, and especially A. Caticha for their many discussions on Entropy, Inference, and Quantum Mechanics.


\begin{thebibliography}{9}
\bibitem{QRE} 
Vanslette, K. Entropic Updating of Probability and Density Matrices. \emph{arXiv}:1710.09373.

\bibitem{book}
Caticha, A. \emph{Entropic Inference and the Foundations of
Physics} (monograph commissioned by the 11th Brazilian Meeting on Bayesian
Statistics -- EBEB-2012); Available online: http://www.albany.edu/physics/ACaticha-EIFP-book.pdf.

\bibitem{Shore1}
Shore, J. E.; Johnson, R. W.; Axiomatic derivation of the Principle of Maximum Entropy and the Principle of Minimum Cross-Entropy. \emph{IEEE Trans. Inf. Theory \bf{1980}} IT-26, 26-37.

\bibitem{Shore2}
Shore, J. E.; Johnson, R. W. Properties of
Cross-Entropy Minimization. \emph{IEEE Trans. Inf. Theory \bf{1981}} IT-27, 472-482.

\bibitem{Csiszar1991}
Csisz\'{a}r, I. Why least squares and maximum entropy: an axiomatic approach to inference for linear inverse problems. \emph{Ann. Stat.\bf{1991}}, \emph{19}, 2032.

\bibitem{Skilling1}
Skilling, J. The Axioms of Maximum Entropy. \emph{Maximum-
Entropy and Bayesian Methods in Science and Engineering}, Dordrecht, Holland, 1988; Erickson, G. J.; Smith, C. R.; Kluwer Academic Publishers: 1988.

\bibitem{Skilling2}
Skilling, J. Classic Maximum Entropy. \emph{Maximum-
Entropy and Bayesian Methods in Science and Engineering}, Dordrecht, Holland, 1988; Skilling, J.; Kluwer Academic Publishers: 1988.

\bibitem{Skilling3}
 Skilling, J. Quantified Maximum Entropy. \emph{Maximum En-
tropy and Bayesian Methods in Science and Engineering}, Dordrecht, Holland, 1988; Foug\'ere, P. F.; Kluwer Academic Publishers: 1990.

\bibitem{Neumann} 
von Neumann, J. \emph{Mathematische Grundlagen der Quantenmechanik}; Springer-Verlag: Berlin, Germany, 1932. [English translation: \emph{Mathematical Foundations of Quantum Mechanics}; Princeton University Press, Princeton NY, USA, 1983.].

\bibitem{Luders} 
 L\"{u}ders, G. \"{U}ber die Zustands\"{a}nderung durch den Messprozess. \emph{Ann. Phys. Leipzig \bf{1951}}, \emph{8}, 322.

\bibitem{Korotkov1} 
 Korotkov, A. Continuous quantum measurement of a double dot. \emph{Phys. Rev. B \bf{1999}}, \emph{60},  5737-5742.

\bibitem{Korotkov2} 
 Korotkov, A. Selective quantum evolution of a qubit state due to continuous measurement. \emph{Phys. Rev. B \bf{2000}}, \emph{63}, 115403, 10.1103/PhysRevB.63.115403.

\bibitem{Jordan} 
Jordan, A.; Korotkov, A. Qubit feedback and control with kicked quantum nondemolition measurements: A quantum Bayesian analysis. \emph{Phys. Rev. B \bf{2006}}, \emph{74}, 085307.

\bibitem{Jacobs} 
Jacobs, K. \emph{Quantum Measurement Theory and its Applications}; Cambridge University Press, 2014. 

\bibitem{Davies} 
Davies, E. \emph{Quantum Theory of Open Systems}; Academic, London, 1976. 

\bibitem{Kraus} 
Kraus, K. \emph{States, Effects, and Operations: Fundamental Notions of Quantum Theory}; Springer, Berlin, 1983. 

\bibitem{Holevo} 
Holevo, A. \emph{Statistical Structure of Quantum Theory}; Springer, Berlin, 2001. 

\bibitem{Nielsen} 
Nielsen, M.; Chuang, I. \emph{Quantum Computation and Quantum Information}; Cambridge University Press, Cambridge, 2000. 

\bibitem{Hellmann} 
Hellmann, F.; Kami\'nski, W.; Kostecki, P. Quantum collapse rules from the maximum relative entropy principle. \emph{New J. Phys. \bf{2016}}, \emph{18}, 013022.

\bibitem{Kostecki} 
Kostecki, R.L\"{u}ders' and quantum Jeffrey's rules as entropic projections. \emph{arXiv}, 1408.3502v1.


\bibitem{Giffin} 
Giffin, A.; Caticha, A. Updating Probabilities. Presented at MaxEnt (2006).  \emph{MaxEnt 2006, the 26th International Workshop on Bayesian Inference and Maximum Entropy Methods}, Paris, France.

\bibitem{GiffinThesis} 
Giffin, A. \emph{Maximum Entropy: The Universal Method for Inference}. University at Albany (SUNY), 2009.

\bibitem{Jaynes1} 
Jaynes, E.T. Information Theory and Statistical Mechanics. \emph{Phys. Rev. \bf{1957}}, \emph{106}, 620-630.

\bibitem{Jaynesbook} 
Jaynes, E.T. \emph{Probability Theory: The Logic of Science}; Cambridge University
Press, Cambridge, UK, 2003.

\bibitem{Jaynes2} 
Jaynes, E.T. Information Theory and Statistical Mechanics II. \emph{Phys. Rev. \bf{1957}}, \emph{108}, 171-190.

\bibitem{Goold} 
Goold, J.; Huber, M.; Riera, A.; del Rio, L.; Skrzypczyk, P. The role of quantum information in thermodynamics -- a topical review. \emph{J. Phys. A: Math. Theor. \bf{2015}}, \emph{49}, 143001.


\bibitem{Schlosshauer} 
Schlosshauer, M. \emph{The quantum-to-classical transition and decoherence}; Springer, Berlin, 2014.

\bibitem{Wooters1} 
Hughston, L.; Jozsa, R.; Wooters, W.  A complete classification of quantum ensembles having a given density matrix. \emph{Physics Letters A \bf{1993}}, \emph{183}, 14-18.

\bibitem{Wooters2} 
Jozsa, R.; Robb, D.; Wooters, W.  A lower bound for accessible information in quantum mechanics. \emph{Phys. Rev. A \bf{1994}}, \emph{49}, 668.

\bibitem{Neumark} 
Neumark, M.  \emph{Izv. Akad. Nauk SSSR, Ser. Mat. \bf{1940}}, \emph{4}, 53, 277.

\bibitem{Gaasbeek} 
Gaasbeek, B. Demystifying the Delayed Choice Experiments. \emph{arXiv}, 1007.3977v1.



\end{thebibliography}
\end{document}